# Quantification of Electrolyte Degradation in Lithium-ion Batteries with Neutron Imaging Techniques

Yonggang Hu[1], Yiqing Liao[1], Lufeng Yang[2,3], Ke Zhang[1], Yufan Peng[1], Shijun Tang[4], Shengxiang Wang[2,3], Meifang Ding[1], Jiahao Wu[1], Jianrong Lin[4], Jinding Liang[5], Yimin Wei[5], Yanting Jin[1], Zhengliang Gong[4], Anatoliy Senyshyn[6], Jie Chen[2,3,*] & Yong Yang[1,4,*]

[1]State Key Laboratory for Physical Chemistry of Solid Surfaces, Collaborative Innovation Center of Chemistry for Energy Materials (iChEM), Innovation Laboratory for Sciences and Technologies of Energy Materials of Fujian Province (IKKEM), Department of Chemistry, College of Chemistry and Chemical Engineering, Xiamen University, Xiamen, 361005, China. [2]Institute of High Energy Physics, Chinese Academy of Sciences (CAS), Beijing 100049, China. [3]Spallation Neutron Source Science Center, Dongguan 523803, China. [4]College of Energy, Xiamen University, Xiamen, 361005, China. [5]Contemporary Amperex Technology Co., Ltd. (CATL), Ningde 352100, China. [6] Heinz Maier-Leibnitz Zentrum (MLZ), Technische Universität München, Lichtenbergstr. 1, 85748, Garching, Germany. [7]These authors contributed equally: Yonggang Hu, Yiqing Liao.

*e-mail: chenjie@ihep.ac.cn; yyang@xmu.edu.cn.

**ABSTRACT:** Non-destructive characterization of lithium-ion batteries (LIBs) provides critical insights for optimizing performance and lifespan while preserving structural integrity. Optimizing electrolyte design in commercial LIBs requires consideration of composition, electrolyte-to-capacity ratio, spatial distribution, and associated degradation pathways. However, existing non-destructive methods for studying electrolyte infiltration, distribution, and degradation in LIBs lack the spatiotemporal resolution required for precise observation and quantification of the electrolyte. In this study, we employ neutron imaging (NI) with sufficient spatial resolution ~150 μm and large field of view 20×20 cm$^2$ to quantitatively resolve the electrolyte inventory (EI) and distribution within LiFePO$_4$/graphite pouch cells under high-temperature accelerated aging. Quantitative standard curves based on neutron transmission attenuation reveal a clear electrolyte dry-out threshold at 3.18 g Ah$^{-1}$ and the two stages evolutions of EI during cell aging were quantified. By integrating non-destructive electrochemical diagnostics, accelerated graphite material loss and liquid phase Li$^+$ diffusion degradation is observed during pore-drying. Further analysis, including operando cyclic aging, reveals that the neutron transmission below the saturation reference is due to the enrichment of hydrogen nuclei within the solid-electrolyte interphase (SEI). Assumed pore-drying does not occur, the SEI signal of the electrodes can be quantitatively decoupled during ageing. Combined analyses with NI, TOF-SIMS, and SEM reveal that high EI cells exhibit uniform SEI growth and reduced degradation, while low EI cells show uneven SEI formation, accelerating capacity loss. This study unveils a dynamic electrolyte infiltration-consumption-dry-out process in LIBs, offering non-destructive and quantitative insights to guide sustainable and durable battery development.

The safe and efficient utilization of clean energy is essential for achieving carbon peaking and carbon neutrality goals[1]. Lithium-ion batteries (LIBs), as key secondary energy storage devices, feature high energy density, safety, and long lifespan, making them indispensable for mitigating the temporal and spatial imbalances in clean energy distribution[2, 3]. Among LIB components, the organic liquid electrolyte—often called the "lifeblood"—fills the intricate pore structures and provides ionic pathways connecting electrodes, enabling localized reactions[4]. Its properties, including ionic conductivity and wettability[5], are critical to battery performance. However, during operation, the electrolyte's state, spatial distribution, quantity, and composition dynamically evolve due to physical, chemical, and electrochemical processes. These changes include phase transitions driven by ambient temperature[6, 7], electrolyte motion caused by the expansion and contraction of electrode active materials[8, 9], lithium salt decomposition at high temperatures[10], and the formation of the solid-electrolyte interphase (SEI) on the negative electrode (NE) surface. SEI formation plays a decisive role in battery stability, but electrolyte-related side reactions contribute to lithium inventory loss (LLI), impacting the thermodynamic capacity[11]. Moreover, degraded electrolyte and SEI affect lithium-ion transport kinetics and interfacial processes, leading to accelerated performance degradation[12].

Electrolyte inventory (EI) is particularly important, as levels below the internal pore volume threshold trigger severe thermodynamic and kinetic degradation[13, 14, 15]. The depletion of EI first leads to ionic transport pathway loss, which subsequently increases local current density, accelerates Li$^+$ depletion and heightens liquid-phase diffusion resistance, ultimately manifesting as a rapid capacity decline. Traditional methods, such as differential weight and solvent extraction, quantify EI destructively and *ex situ* at specific states of health (SoH) or charge (SoC)[16]. These approaches, however, cannot dynamically track EI changes or spatial distribution, limiting their relevance for operational studies.

Advancements have introduced non-destructive diagnostic techniques to address these limitations[6, 7, 17, 18]. Differential thermal analysis (DTA) links electrolyte components to thermal peaks during controlled heating[6]. Despite its non-destructive nature, DTA requires freezing the battery, precluding *in situ* applications. Furthermore, its quantitative accuracy is compromised by heat conduction delay and limited spatial resolution. Spatially resolved neutron powder diffraction (NPD) has provided critical insights by monitoring electrolyte peaks across battery cross-sections at cryogenic temperatures (150 K)[7]. However, NPD remains constrained by long acquisition times, relative rather than absolute quantification due to frozen electrolyte complexities and limited potential for *in situ* studies. Ultrasound imaging combined with energy dispersive spectroscopy (EDS) offers spatially resolved electrolyte quantification in pouch cells with a pixel size of 150×150 μm[17]. While promising for *in situ* diagnostics, the technique faces critical challenges: ultrasound signals attenuate exponentially in gas phases, reducing effectiveness in low EI cells; additionally, the proposed thickness correction method becomes impractical in real-world batteries constrained by mechanical housings. Neutron imaging (NI) has emerged as a promising tool for investigating electrolyte infiltration[19, 20], motion[18, 21], consumption[18], and associated gas evolution[18, 22], due to the high neutron attenuation coefficients of hydrogen (H) and lithium (Li) in the electrolyte[23]. While prior applications of NI have revealed dynamic changes in electrolyte behavior during charge-discharge cycles[18, 21] and aging[18], these studies have largely been qualitative, with limited quantitative insights into EI dynamics. This highlights the need for advanced methodologies to leverage NI's non-destructive and *in situ* capabilities for more precise EI characterization. These studies collectively highlight advancements in electrolyte characterization but underscore persistent challenges in achieving accurate, non-destructive, and direct observation of EI dynamics and spatial distribution under operational conditions. We developed an advanced diagnostic framework to overcome these limitations targeting non-destructive measurement, *in situ* applicability, quantitative accuracy, and high spatial resolution. Using 110 mAh LiFePO$_4$ (LFP)/graphite (Gr) pouch cells with two initial electrolyte levels, we investigated the thermodynamic and kinetic performance degradation during 65°C accelerated aging, focusing on the electrolyte degradation behavior. Non-destructive electrochemical diagnostics revealed significant graphite material loss and a marked decline in Li$^+$ diffusion during electrolyte dry-out. To identify the root causes of diffusion impairment, we employed energy-resolved neutron imaging instrument (ERNI)[24] at the China Spallation Neutron Source (CSNS), achieving ~150 μm resolution (50×50 μm pixel size) and a 20×20 cm² large field of view with single-image exposure time 20 s. Quantitative standard curves revealed a clear transition between the two stages of electrolyte inventory loss (EIL): excessive electrolyte inventory loss (ExEIL) and pore-drying electrolyte inventory loss (PDEIL). Neutron images of aged cells confirmed the EI at different states of health (SoH), with validation through differential weight showing errors below 26.6 mg during ExEIL and 43.1 mg during PDEIL. A systematic deviation in NI-based electrolyte quantification within the electrode region was attributed to hydrogen nuclei enrichment in the SEI, as confirmed by additional experiments including *operando* cyclic aging. The SEI signal can be quantitatively decoupled during ExEIL. Combined analyses with NI, TOF-SIMS and SEM reveal the correlation between EIL, SEI growth, and irreversible capacity degradation. This study provides a comprehensive understanding of the electrolyte infiltration–consumption–dry-out process in LIBs under aging conditions and offers a non-destructive, quantitative framework for studying electrolyte behavior with sufficient spatial and temporal resolution.

**Results**
**Accelerated aging and electrochemical diagnosis for electrolyte inventory loss**
Under conventional mild aging conditions, EIL in well-designed cells is a gradual process, and observing the PDEIL, commonly referred to as "dry-out"[12], typically requires prolonged testing to achieve a deeply aged state. The substantial economic costs (e.g., battery testing resources) and time investment make it challenging to decouple the effect of EI on cell aging performance. To accelerate and effectively simulate the two stages of EIL, we designed experiments based on two strategies: (1) controlling the initial electrolyte injection volume and (2) selecting high-temperature accelerated aging conditions to expedite EIL. The choice of electrolyte injection volume was referenced to the cell's intrinsic pore volume $V_{pore,cell}$, which was calculated using the design parameters of the electrodes and separator:

$$V_{pore,cell} = V_{pore,PE} + V_{pore,sep} + V_{pore,NE}$$
$$= 2 \times A_{PE} \cdot L_{PE} \cdot \varepsilon_{PE} + 2 \times A_{sep} \cdot L_{sep} \cdot \varepsilon_{sep} + 2 \times A_{PE} \cdot L_{PE} \cdot \varepsilon_{PE} \approx 225 \mu L \quad (1)$$

Where $V_{pore,x}$ represents the pore volume of component $x$ (positive electrode (PE), separator (sep), and NE). $A_x$ is the geometric area of component $x$. $L_x$ is the thickness of component $x$ (i.e., single-layer coated electrode thickness or separator thickness). $\varepsilon_x$ is the porosity of component $x$. $V_{pore,cell}$ was calculated to be approximately 225 μL (see Supplementary Table 1 for detailed parameters). As described in "Methods" section, to cover the individual cell variations and allow cells at the fresh state (corresponding to 100% state of health (SoH) after formation, or beginning of life (BoL)) to approach but not enter the PDEIL stage, an initial electrolyte injection volume of 300 μL was selected for the low initial EI group (labeled as the "low EI group"). For the high initial EI group, 400 μL was chosen to prevent significant dry-out even at the defined end of life (EoL, 70% SoH), labeled as the "high EI group". To further accelerate EIL, we subjected both low and high EI groups to high-temperature (65°C), 1C charge/discharge cycling, which simultaneously promotes chemical self-decomposition and electrochemical side reactions of the electrolyte components. The tested cell type and fixtures are shown in Supplementary Fig. 1.

As shown in Fig. 1a, capacity fade at elevated temperatures can be categorized into three distinct stages based on the fading rate. In stage I (100–95% SoH), the cells exhibit a relatively slow capacity fade. In stage II (95–77% SoH), the fade accelerates significantly, while in stage III (77–70% SoH), the decline transitions into a more gradual plateau. Notably, while dry-out was observed, as shown in subsequent sections, no capacity drop-off was detected throughout the degradation process. Using accurate EMF curves obtained via extrapolated electromotive force test (Extrap-EMF) diagnostics (detailed description and protocols can be seen in the "Methods" section and Supplementary Table 2), we quantitatively identified key thermodynamic degradation modes (TDMs) through the previously developed degradation mode analysis (DMA) program[25]. From the results of LLI shown in Fig. 1b, the aging trend of LLI aligns closely with the 1C capacity fade, indicating that LLI is the dominant degradation mode. In contrast, Fig. 1c reveals that the loss of active material in the negative electrode ($LAM_{NE}$) exhibits a different trend from LLI. The distinct trends between the two groups suggest differing dependencies of $LAM_{NE}$ on degradation variables (including cycles, LLI, and time). For the high EI group, $LAM_{NE}$ increases almost linearly, reaching 8.42% at EoL. This linear dependency suggests that the $LAM_{NE}$ is caused by mechanically-induced particle cracking, which has a cycle number dependency[26]. However, in the low EI group, $LAM_{NE}$ exhibits a two-phase trend, with an initially rapid increase followed by a slower rate, ultimately reaching 11.35% at EoL - 1.35 times higher than that of the high EI group. These results suggest that an initial EI only slightly above the pore volume leads to significantly faster capacity fade and LLI rates during stages I and II, as well as a larger $LAM_{NE}$ at the same SoH. Fig. 1d further shows the relationship between $LAM_{NE}$ and SoH (and thus LLI). In the low EI group, $LAM_{NE}$ exhibits a linear correlation with SoH, indicating that $LAM_{NE}$ is directly associated with the extent of irreversible electrochemical side reactions responsible for capacity loss.

The kinetic degradation of the cells was analyzed using direct current resistance (DCR) obtained from short relaxation time galvanostatic intermittent titration technique (Short-Rest-GITT) diagnostics at the aging temperature and temperature-varied electrochemical impedance spectroscopy (TV-EIS) conducted at 50% SoC (TV-EIS@50%SoC). The testing details are provided in "Methods". The 1s DCR (Fig. 1e) captures polarization changes in fast kinetic processes characterized by short time constants, which are generally associated with variations in Ohmic resistance, SEI resistance, and charge-transfer impedance. In contrast, the 30s DCR (Fig. 1f) reflects polarization changes in slower kinetic processes with larger time constants, primarily representing liquid-phase $Li^+$ diffusion impedance. The similar trends in 1s DCR for both groups indicate comparable polarizations at the electrochemical reaction interfaces under the aging temperature. However, significant differences were observed in 30s DCR between the two groups after ~90% SoH, where the low EI group exhibited a markedly faster increase. By EoL, the 30s DCR of the low EI group was approximately 13% higher than that of the high EI group, suggesting that electrolyte consumption during aging led to a deterioration in liquid-phase Li⁺ diffusion performance at the aging temperature. Another essential tool for kinetic analysis is TV-EIS@50%SoC, which effectively decouples various kinetic processes across the frequency spectrum. The Nyquist plots of TV-EIS at 50% SoC (Supplementary Fig. 2 and Supplementary Fig. 3) reveal a progressively longer high-frequency region with a 45-degree sloped line as the test temperature decreases. According to the transmission line model (TLM), the length of this line segment's projection onto the real axis corresponds to the ionic resistance in the pores within the electrode regions $R_{ion}$[27]. As the EIS test temperature decreases, $R_{ion}$ becomes more distinguishable from the charge transfer resistance of lithium intercalation $R_{ct}$ due to differences in activation energy. Therefore, we selected $R_{ion}$ at -20°C (the lowest EIS temperature in this study) to illustrate the differences in pore resistance across different cells. As shown in Fig. 1g, $R_{ion}$ at -20°C in the low EI group increases significantly after aging below 90% SoH, reaching 3.48 times that of the high EI group at EoL. This indicates severe electrolyte depletion within the electrode pores (not in the separator pores), consistent with the PDEIL stage. Although $R_{ion}$ at -20°C cannot definitively resolve the contributions of pore dry-out between the PE and NE, the larger pore size of graphite particles in the NE makes them more susceptible to dry-out. This is further supported by the pronounced $LAM_{NE}$ increase observed in the low EI group, suggesting substantial pore dry-out in the NE. Notably, while significant pore dry-out was observed at low temperatures, it did not result in severe kinetic degradation at high temperatures. This indicates that elevated temperatures largely compensate for or mask the kinetic degradation, particularly ensuring unabated charge-transfer reaction kinetics.

Based on the kinetic parameters, we infer that the Low EI group transitions from the ExEIL to the PDEIL stage around 90% SoH. However, this does not explain the significantly faster degradation rate observed in the Low EI group during capacity fading stage I, where no substantial drying occurred. *Post mortem* electrolyte component quantification results (see "Methods" section and Supplementary Table 3) reveal that although the mass ratio of major electrolyte components, including lithium salts and solvents, to total electrolyte mass, is nearly identical between the two groups at BoL, there is a significant difference in the mass ratio of the additive VC. The VC content in the High EI group is approximately 3.5 times higher than in the Low EI group at BoL. During the early aging stage, VC is preferentially reduced at the NE surfaces because its reduction potential is higher compared to other species. The lower VC concentration in the Low EI group leads to a greater participation of lithium salts and solvents in electrochemical side reactions, accelerating the degradation rate. The depletion of VC components at 90% SoH explains the accelerated aging observed in both groups during capacity fading stage II. However, the significantly faster degradation rate in the Low EI group during this stage suggests that the PDEIL stage introduces additional acceleration of degradation, which will be further discussed in the final section of the "Results" chapter. The reasons behind the slowdown in degradation rate during Stage III remain unclear. Still, we hypothesize that this is due to a reduction in single-cycle charge throughput caused by 1C capacity loss, leading to a decreased rate of single-cycle capacity degradation.

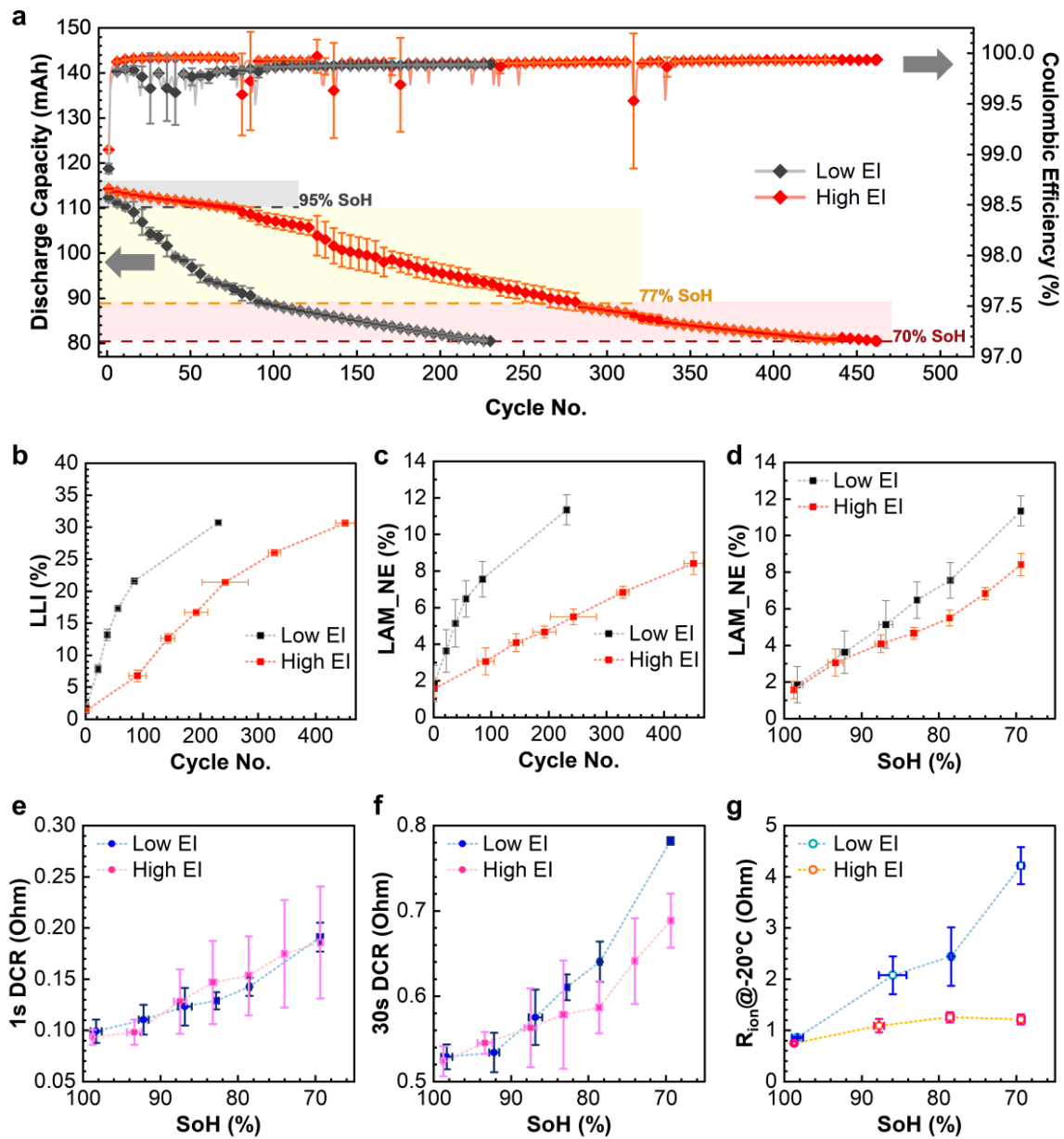

**Fig. 1 | Accelerated aging and electrochemical non-invasive diagnosis. a**, Cyclic aging performance of the different initial EI cells with a charge/discharge rate of 1 C/1 C at 65°C. **b**, Loss of Lithium Inventory (LLI) vs. cycle number. **c**, $LAM_{NE}$ vs. cycle number. **d**, $LAM_{NE}$ vs. SoH. **e**, **f**, 1s (**e**) and 30s (**f**) DCR values calculated from Short-Rest-GITT during aging. It should be noted that the DCR values at specific SoH in the plots represent the averages of DCR values obtained across all SoCs during the corresponding Short-Rest-GITT measurement. **g**, $R_{ion}$ calculated from EIS measurements at -20°C using the transmission line model during aging. It should be noted that at each target SoH of 90%, 80%, and 70%, three cells from each test group were removed from the test channels for subsequent end of aging diagnosis and NI experiments. In contrast, the remaining cells continued aging to deeper aging states. Consequently, in the

100–90% SoH range, all DC electrochemical data and their error bars represent nine samples; in the 90–80% SoH range, six samples; and in the 80–70% SoH range, three samples. One cell from the Low EI group, expected to represent the 90% SoH sample, was accidentally short-circuited during testing, and the EIS data for one High EI group sample at 90% SoH were lost. These issues will not be repeated in the following discussion. Since TV-EIS measurements were performed before aging and after final aging, the EIS data and error bars are based on nine cells at BoL and three cells at other SoHs. Each data point reflects the mean across the relevant samples, and the error bars represent their standard deviation (this statistical approach applies to all figures hereafter).

**Constructing EI quantitative reference base with neutron imaging**

A reference base in the form of quantitative standard curves was established using neutron imaging (NI) technology to determine the EI in aged cells. The performance of the NI system and the validity of image post-processing significantly influence the accuracy of the quantification. Fig. 2a illustrates the key components of the NI system used in this study. The coupled hydrogen moderator (CHM) is chosen as the neutron source for ERNI. The cold and thermal neutrons from CHM are delivered to the cell samples with fixtures (under a pressure of 1.5 MPa) by a neutron guide and collimation system. Transmitted neutrons hit the neutron-sensitive scintillation screen, enabling imaging of the neutron attenuated object. To improve experimental efficiency, a large field-of-view (20 × 20 cm², sampled to 2048 × 2048 pixels) configuration was used, allowing simultaneous imaging of two cell samples. The detailed setup parameters of the neutron imaging system are provided in the "Methods" section. Sensor pixel does not reflect the actual physical resolution. The real resolution was determined using Siemens stars to be in the range of 150 × 150 μm (see Supplementary Fig. 4). The original neutron images are superimposed by the multiple background signals, including interferences from gaskets and fixtures, requiring certain post-processing (Fig. 2b). A detailed description of the post-processing applied to the original images is provided in Supplementary Note 1. The resulting high-quality neutron images contain only the sample information, enhancing comparability between samples and providing a robust quantitative foundation for image-based studies.

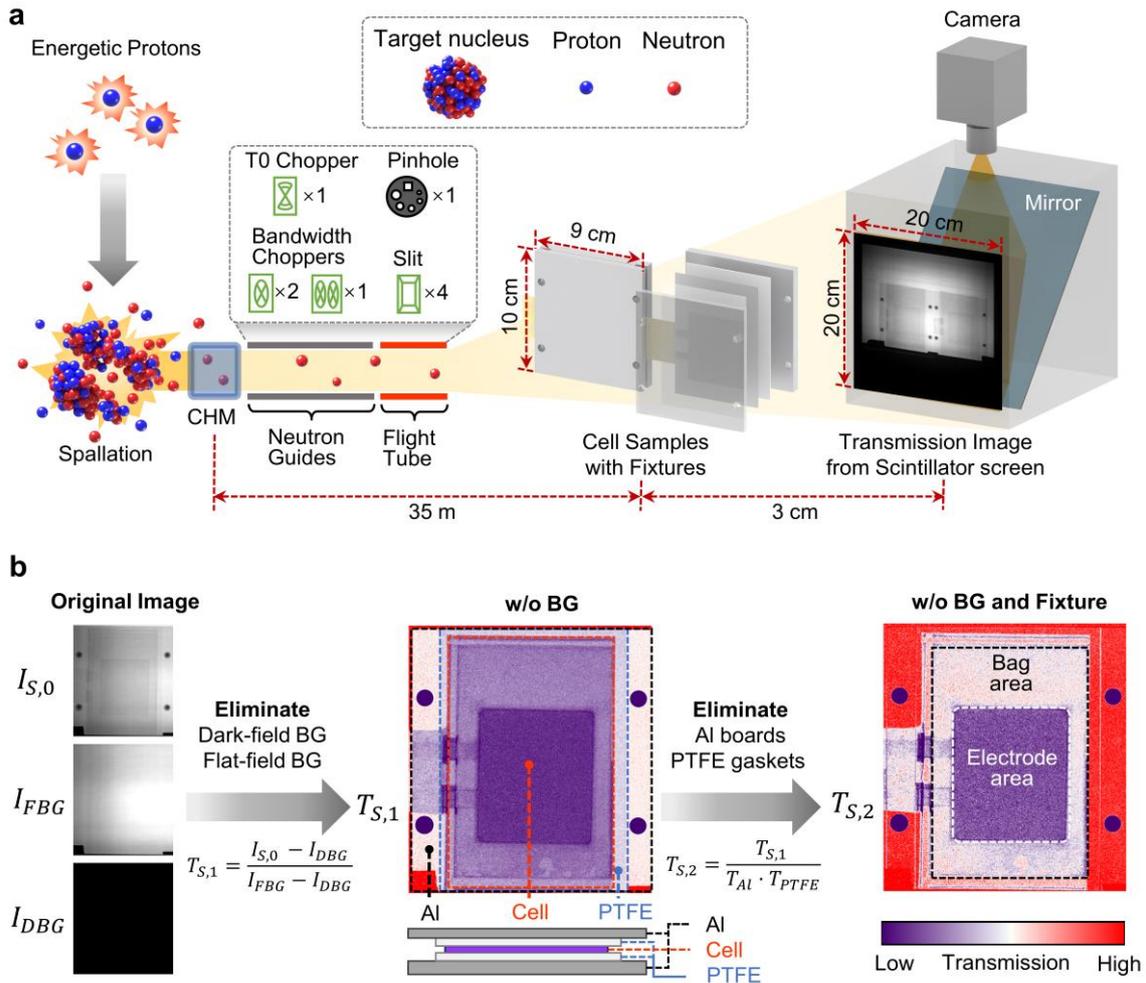

**Fig. 2 | Schematic illustration of the neutron radiography experiment. a**, NI system setup. **b**, The post-processing workflow for the acquired images.

A quantitative reference base standard curve was constructed using a set of standard cells with controlled electrolyte injection to further establish the quantitative relationship between NI signals and EI. The standard cells consisted of 11 dry cells injected with varying amounts of electrolyte: 0.00 g, 0.10 g, 0.20 g, 0.25 g, 0.30 g, 0.35 g, 0.40 g, 0.45 g, 0.50 g, 0.55 g, and 0.60 g. Neutron images of these standard cells were acquired (Fig. 3a, 3b), and quantitative standard curves of the average transmission in the gas bag and electrode regions were obtained as a function of EI (Fig. 3c, 3d). We observed the non-uniform distribution of the electrolyte. Regarding the spatial distribution of the electrolyte in the gas bag, gravity does indeed influence its distribution. The excess electrolyte we observed is primarily located beneath the gas bag, closer to the side facing the Earth's center. In contrast, the neutron transmission signal distribution in the electrode region showed no significant gravity-related pattern, indicating that capillary forces within the micron-scale pores overcame the influence of gravity. Additionally, weak non-uniformities in neutron attenuation were detected, which were attributed to local inhomogeneities. Since the standard cells were not subjected to formation, the EI signals observed in the neutron images were maximally decoupled from other potential influences. We quantified the data by calculating the average transmission within selected regions. A detailed rationale for this approach is provided in Supplementary Note 2. The standard curves revealed distinct trends: as EI increased from 0.00 g, the average transmission in the electrode region decreased linearly, while the transmission in the gas bag region remained nearly constant 93.8%, corresponding to the preferential filling of the pore space. However, when EI exceeded 3.18 g Ah$^{-1}$, the average transmission in the electrode region stabilized at 87.9%, and a linear decrease in the gas bag transmission was observed, indicating the accumulation of excess electrolyte in the gas bag region after full pore saturation. These results identify a full saturation/dry-out threshold at 3.18 g Ah$^{-1}$, representing a clear transition threshold between the ExEIL stage and the PDEIL stage.

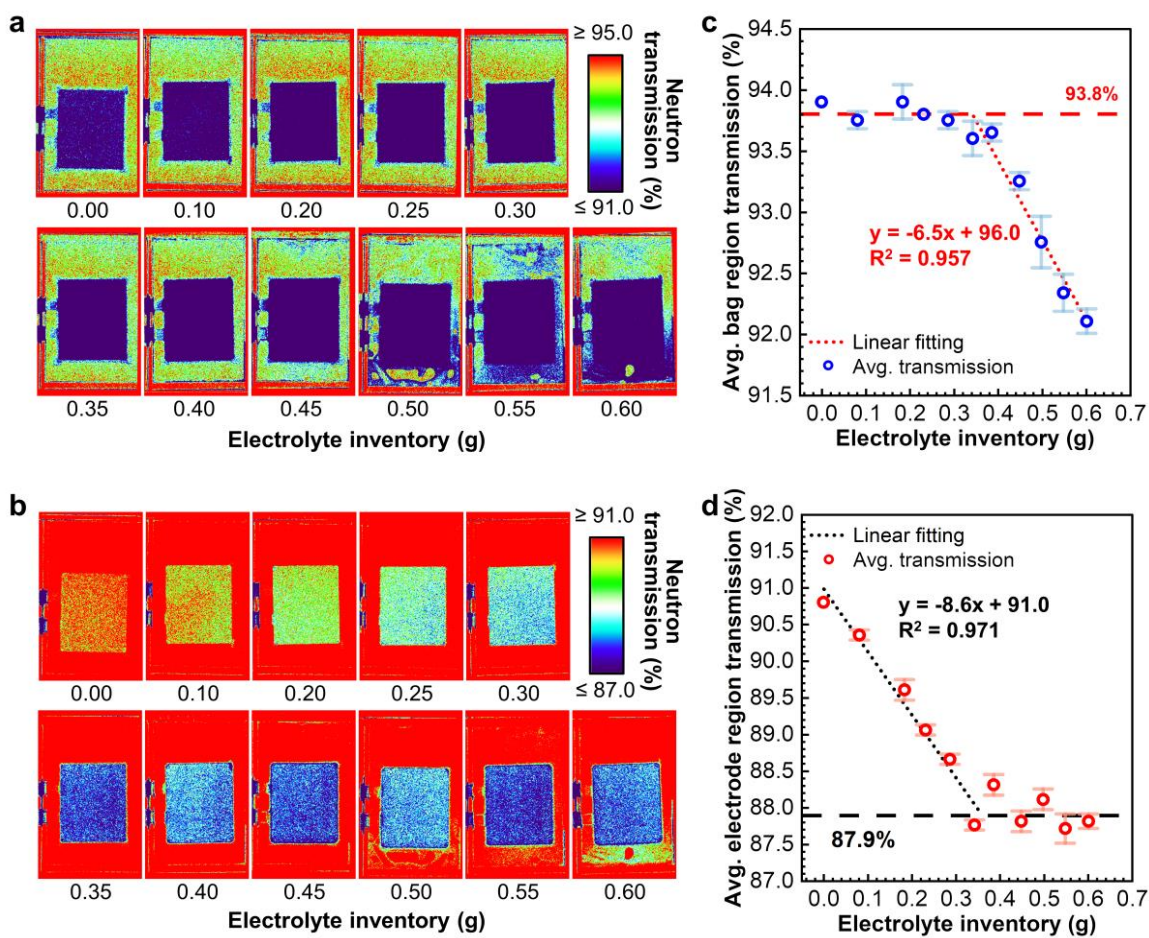

**Fig. 3 | Quantitative standard curve construction for EI quantification in pouch cells via NI. a**, Neutron transmission image with enhanced contrast in the bag region. **b**, Neutron transmission image with enhanced contrast in the electrode region. **c**, Standard curve for the bag region. **d**, Standard curve for the electrode region. Note that the error bars in the standard curves reflect two separate imaging captures of a single sample. The data values in **c** and **d** are provided in Supplementary Table 4.

## Application and validation of NI for EI quantification in aged cells

Using the standard curves established in the previous section, we quantitatively analyzed the changes in EI of aged cells under accelerated aging conditions at 65°C. After cycling to the target SoHs, the aged cells were discharged to 2.5V, and NI experiments were conducted to quantify the EI. The samples consisted of both initial EI groups (low and high), each analyzed at four states of health (SoH: 100%, 90%, 80%, and 70%), with three different cell samples per SoH. As shown in Fig. 4a, changes in transmission in the gas bag region were distinct between the two groups. In the low EI group, no observable electrolyte accumulation was present in the gas bag region from the initial state, and its average transmission remained consistently high throughout aging (Fig. 4c). In contrast, the high EI group initially exhibited clearly identifiable excess electrolyte in the gas bag region, which gradually diminished during aging. Correspondingly, the average transmission in the gas bag region increased progressively (Fig. 4c), indicating a reduction in total EI. The quantitative results from the gas bag standard curve reveal that the electrolyte loss rate in the high EI group was 3.47 mg/%SoH (Fig. 4e).

The transmission trends in the electrode regions (Figs. 4b and 4d) exhibited opposing behaviors for the two initial EI groups: an increase in the low EI group and a decrease in the high EI group, with changes exceeding the measurement error margin (0.2%). The significant increase in transmission in the low EI group occurred after 90% SoH. Coupled with sharp increases in key thermodynamic and kinetic degradation factors at 90% SoH, including $LAM_{NE}$, 30s DCR, and $R_{ion}$ @-20°C, this phenomenon can be attributed to pore dry-out in the electrode region, marking the onset of the PDEIL stage. This also explains the linear relationship observed between $LAM_{NE}$ and SoH in the low EI group: the irreversible electrochemical side reactions causing capacity loss are directly proportional to the extent of PDEIL, with greater PDEIL leading to higher $LAM_{NE}$. Therefore, $LAM_{NE}$ in the low EI group can be considered primarily driven by EIL. At this stage, the EI of the low EI group could not be quantified using the gas bag standard curve but was effectively measured using the electrode region standard curve, yielding an EIL rate of 2.47 mg/%SoH, which is 0.71 times that of the high EI group. This result appears counterintuitive—despite a faster capacity fade rate, the EIL rate in the low EI group is lower than that in the high EI group. However, it is important to recognize that EIL is not solely attributed to electrochemical side reactions contributing to capacity fade but also includes time-dependent chemical self-decomposition of the Li salt. When comparing the low EI group to the high EI group at the same SoH, the low EI group exhibits a shorter cycling duration due to its accelerated degradation rate, resulting in relatively less high-temperature decomposition of the Li salt.

We conducted destructive differential weight analysis on two samples per condition following NI experiments to validate the accuracy of NI-based EI quantification. The detailed procedure for this analysis is described in the "Methods" section. As shown in the lower panel of Fig. 4f, the EI results obtained from NI and differential weight exhibited consistent trends, confirming the reliability of the NI-based observations, particularly the lower EIL rate in the low EI group. The upper panel of Fig. 4f shows that the maximum absolute errors for EI quantification were 26.6 mg for the high EI group and 43.1 mg for the low EI group.

We observed a notable result during the error analysis: the EI values quantified using the electrode region standard curve for the low EI group consistently exhibited a positive error. A potentially related yet more perplexing observation is the decrease in transmission within the electrode region for the high EI group, as shown in Fig. 4d. During the construction of the standard curve, we identified a significant trend: the transmission in the electrode region stabilizes at 87.9% under fully saturated conditions. If transmission solely depended on EI, a value below this stable threshold would imply an increase in the electrode pore volume to accommodate additional electrolyte. However, such an interpretation is unreasonable, given the confinement imposed by the 1.5 MPa clamping pressure. It is worth noting that both of these anomalous results are associated with imaging outcomes in the electrode region. The next section will provide further experimental evidence to explain these anomalies.

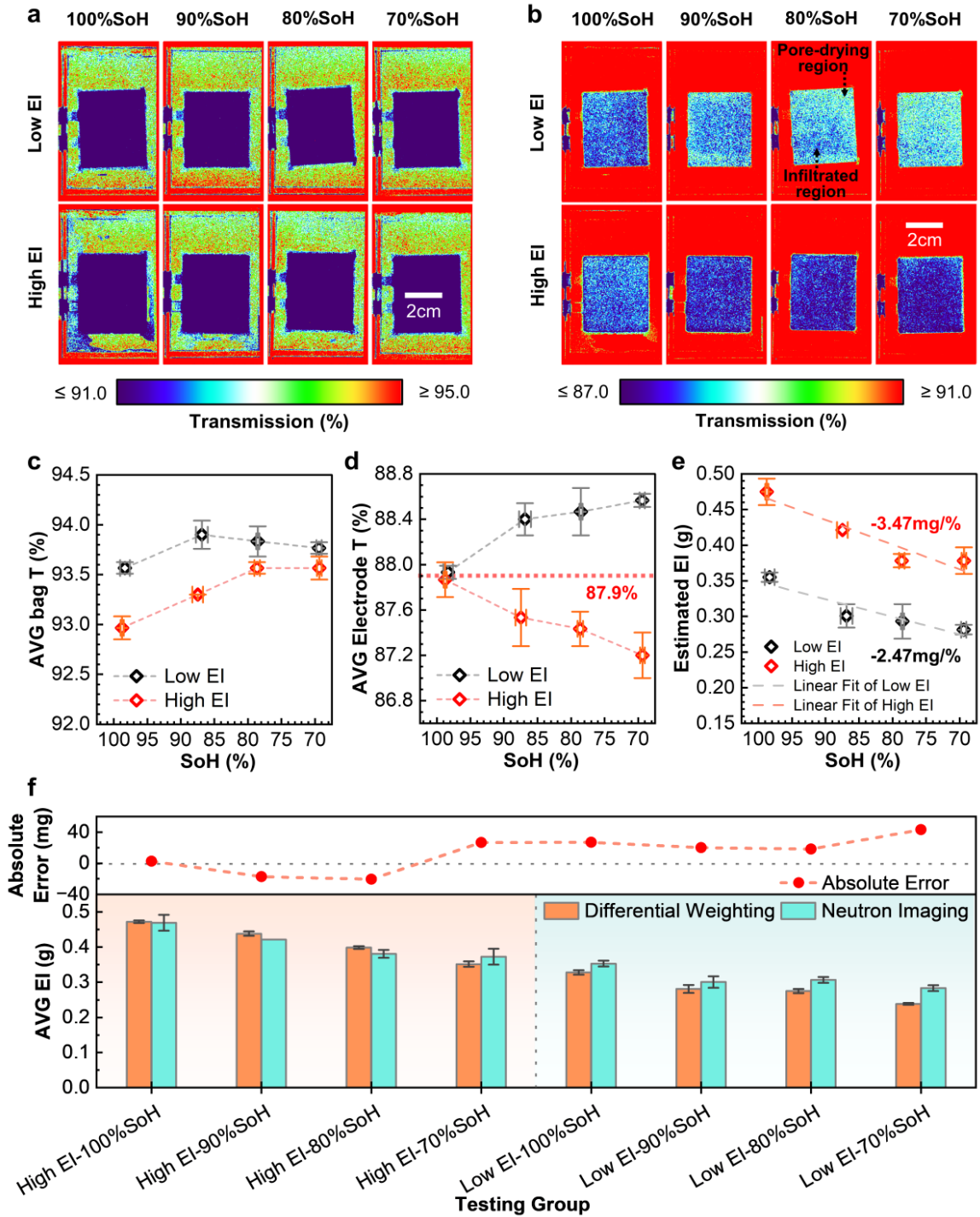

**Fig. 4 | Neutron images and EI quantification results of aged cells. a**, Neutron transmission image with enhanced contrast in the bag region. **b**, Neutron transmission image with enhanced contrast in the electrode region. **c**, **d**, Averaged neutron transmission in the bag (**c**) and electrode (**d**) regions of the aged cells. **e**, Estimated EI of the aged cells. **f**, Comparison of quantitative results between differential weight method and NI. Note that for each test group at each SoH, the NI error bars were derived from three parallel samples. In contrast, the differential weight error bars were obtained from two of those three

samples, as the remaining sample was set aside for other characterization tests.

**Tracking SEI growth during aging**

Fig. 5a compares the neutron transmission signals in the electrode regions of fully infiltrated samples and high EI group samples at varying SoH. The transmission signals are normalized to a fully infiltrated reference value of 87.9%, with deviations shown relative to this reference. At BoL, the high EI group exhibits transmission signals nearly identical to unformed fully infiltrated samples. However, the transmission progressively decreases during aging, falling below the reference value.

To understand this trend, we comprehensively considered all potential changes occurring in the fully infiltrated electrode region during battery aging. A reasonable initial hypothesis assumes that transmission changes are directly linked to gas, liquid, and solid phase variations. Gas-phase formation, attributed to chemical and electrochemical degradation of the electrolyte, is unlikely to lower transmission, as neutron-sensitive nuclei in gas phases are typically sparse. Additionally, gas formation would displace the liquid phase from pores, increasing transmission—contrary to the observed trend. This suggests that gas-phase effects can be excluded. The presence of excess EI in the gas bag region (Fig. 4a) further supports the hypothesis that the high EI group maintains sufficient electrolyte in the electrode region during aging, preventing significant pore drying despite possible minor gas bubble blockage. The electrolyte liquid phase changes are substantial during aging, as the $LiPF_6$, EC, EMC, and VC components evolve. To assess the impact of these changes on neutron transmission, NI was performed on cells filled with electrolytes of varying lithium salt concentrations (0–1.5 M $LiPF_6$) and solvent ratios (EC:EMC = 2:8 to 4:6). As shown in Figs. 5b and 5c, neither lithium salt concentration nor solvent ratio induced transmission changes beyond the error margin (0.2%). This stability is attributed to the negligible density changes in the electrolyte, preserving the spatial distribution of hydrogen nuclei, the primary contributors to neutron attenuation. Additional control experiments altering fixture pressure (Supplementary Fig. 5) confirm that minor pressure variations, simulating stress from electrode volume expansion, also have negligible impact on transmission. Notably, irreversible electrode expansion would reduce pore electrolyte volume and increase transmission, contradicting observations. While irreversible expansion is inevitable during aging, accelerated aging at elevated temperatures minimizes stress cycles and enhances Li$^+$ diffusion, mitigating its impact on pore volume. Solid-phase changes during aging primarily involve the degradation of active/inactive electrode materials and SEI growth. At 2.5V, ~100% of the active lithium inventory ($LI_{active}$) is in the PE. Stoichiometry shifts due to LLI alter this distribution, but the conservation of total lithium nuclei remains:

$$LI_{active,initial} - LI_{active,aged} = LI_{inactive,aged} = LLI \qquad (2)$$

and then

$$LI_{active,initial} = LI_{active,aged} + LI_{inactive,aged} \qquad (3)$$

Here, $LI_{active,initial}$ and $LI_{active,aged}$ represent the $LI_{active}$ within all electrode materials at the BoL and after aging, respectively. $LI_{inactive,aged}$ denotes the inactive lithium inventory in the aged cell, which typically includes lithium trapped in the SEI or in electrode materials that can no longer participate in electrochemical reactions. This conservation relationship indicates that, regardless of the extent of aging, the total lithium nuclei in the solid phase of the electrode region remain constant. The *operando* NI experiments (65°C 2C/2C cyclic accelerated aging) during the lithiation process further support this conclusion (The description of the experimental setup can be found in the "Methods" section and results in Supplementary Video 1 and Supplementary Fig. 7). After formation and 3 cycles before stabilized aging, we observed <0.1% changes in neutron transmission in the electrode region within one charge/discharge process, effectively ruling out the possibility that differences in the physical state of lithium nuclei (e.g., embedded in the PE or NE) significantly contribute to variations in neutron attenuation coefficients. Surprisingly, the *operando* aging experiment revealed periodic transmission variations in both the electrode region and the gas bag region (Supplementary Fig. 7a), closely correlated with the voltage profile despite their relatively small amplitudes (<0.1%). Meanwhile, as shown in Supplementary Fig. 7b, when we observed the averaged transmission across the entire cell region (gas bag + electrode), these periodic oscillations nearly disappeared, leaving only a trend close to baseline evolution. This observation indicates that *operando* NI successfully captured the motion of electrolyte being absorbed into and expelled from the electrode region during cycling, driven by the expansion and contraction of electrode active materials[28]. A notable trend observed during the initial three cycles is the significant decrease (~0.4%) in transmission within the electrode region, accompanied by a corresponding increase (~0.24%) in transmission within the gas bag region. Thereafter, transmission in both regions continues to evolve at a slower rate, maintaining the same directional trends. The pronounced overall trend in transmission changes exceeds the magnitude expected from the expansion and contraction of the electrode, making it insufficient to attribute this entirely to irreversible electrode layer expansion.

Based on the systematic exclusion of all potential factors influencing neutron transmission in the electrode region, combined with the results of *operando* aging experiments, the results point to a single plausible explanation: the decrease in neutron transmission is associated with the growth of the SEI. This explanation also accounts for the observed baseline evolution in Supplementary Figs. 7a and 7b. Although the total amount of neutron-sensitive elemental nuclei in the pouch remains unchanged, their aggregation forms may alter due to electrochemical reactions, and these changes in aggregation can, to some extent, influence the neutron transmission. Fig. 5d illustrates the SEI-related signal, derived by subtracting the fully infiltrated reference value from the neutron transmission of the electrode region in the high EI group, alongside the time required for specific fragments to sputter to stability ($t_{stable}$) as determined by TOF-SIMS. The $t_{stable}$ values of $C_6^-$, $CH_3O^-$, and $LiO^-$ fragments were selected to represent the overall SEI, the organic SEI layer, and the inorganic SEI layer thicknesses, respectively. TOF-SIMS results show continuous growth of the SEI layer during aging, although the growth rate slows over time. Examination of the organic and inorganic SEI layers reveals consistent thickening, with the inorganic layer's growth rate decreasing while the organic layer's thickness increasing. Notably, the growth trend of the organic layer corresponds closely with changes in the SEI-related signal observed via NI. This suggests that the organic SEI layer, rich in hydrogen, dominates the influence on NI signals due to the high neutron attenuation of hydrogen compared to lithium.

Building on these observations, Fig. 5e and 5f summarize the relationship between neutron transmission and SEI growth in the electrode region. Fig. 5e depicts the pre-aging state of the high EI group cell, where the gas bag region contains excess EI and all $LI_{active}$ resides in the cathode, resulting in higher neutron transmission compared to aged cells. In contrast, Fig. 5f shows EoL state, where excess EI has been largely depleted, and a significant portion of the initial $LI_{active}$ has been converted into $LI_{inactive}$ within the SEI layer. Continuous electrochemical reduction of the electrolyte replenishes EI depletion in the electrode region, driving SEI growth. The enrichment of hydrogen in the organic SEI layer, rather than lithium, ultimately leads to decreased neutron transmission in the electrode region. It is important to note that the SEI layer, with a thickness on the order of tens of nanometers[29], occupies a negligible volume relative to the micrometer-scale pore volume and thus does not significantly affect the overall pore volume. The proposed mechanism also accounts for the systematic overestimation of EI quantification in the low EI group based on electrode region NI transmission (Fig. 4f). SEI growth in aged cells causes a systematic decrease in transmission, which translates the reduced transmission into an overestimated EI value when directly interpreted using the standard curve.

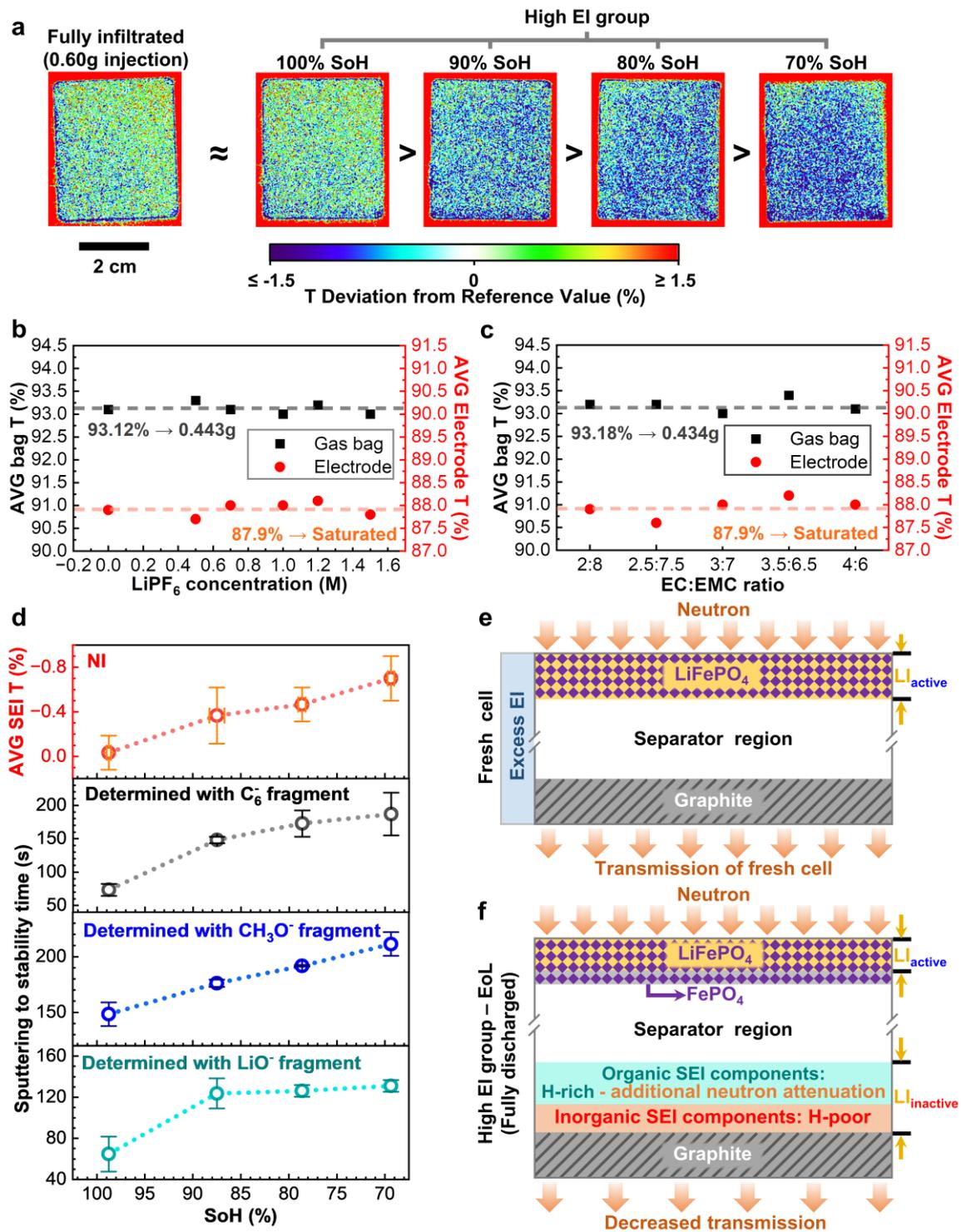

**Fig. 5 | SEI component signals captured by NI. a**, Comparison of the transmission deviation (from the fully infiltrated reference value) between unformed fully infiltrated cells and cells with different SoH in the High EI group. The inequality signs compare the average transmission deviation values of the electrode region. **b,c**, Neutron transmissions of cells with varying LiPF$_6$ concentration (**b**) and EC:EMC ratio (**c**). **d**, SEI signal changes captured by NI compared to the time of specific fragments sputtering to stability obtained with TOF-SIMS, which corresponds to the thickness distribution of the fragment-generating

components. "Sputtering to stability" is defined as the point where the absolute rate of change in the background-corrected normalized intensity of sputtered secondary ion fragments is less than 0.001 a.u./s. Note that the NI error bars were derived from three parallel samples, whereas the TOF-SIMS error bars reflect data collected from two distinct regions of a single electrode sample. **e**, Schematic illustration of the neutron transmission process in a fresh cell. **f**, Schematic illustration of the neutron transmission process at the end of life (EoL) for cells in the High EI group.

**Analysis of aging trend in different initial EI groups**

Our previous work has established that LLI is the dominant degradation mode for LFP/Gr cells. The distinct aging trends observed in the LLI evolution curves of the two initial EI groups (Fig. 1b) suggest significant differences in the associated side reactions and SEI byproducts. Fig. 6a shows the SEM images of graphite NE at BoL, while Figs. 6b and 6c present images at EoL for the two groups. The complete optical and SEM images of samples at BoL and aged states can be seen in Supplementary Fig. 8. Apparent differences in surface morphology are observed across different SoH, with the right panels showing 3D rendered depth distributions of $C_6^-$ fragments, corresponding to the intensity-sputter time profiles in Fig. 6d. At BoL, despite the initial SEI formed during formation, prominent "pit-like" defects are visible on the graphite particle surfaces. The corresponding 3D depth distribution of $C_6^-$ fragments (Fig. 6a, right) reveals that the SEI remains very thin at this stage. For the high EI group at EoL, such "pit-like" defects are no longer evident, and the 3D depth distribution (Fig. 6b, right) shows a uniformly distributed SEI with nanoscale thickness. This suggests that ~30% LLI led to the formation of a relatively thick SEI, effectively covering and filling surface defects. Interestingly, the low EI group at EoL, despite exhibiting the same LLI as the high EI group, shows a distinctly different surface morphology. SEM images consistently reveal numerous low-contrast "pits" on the graphite surface, resembling the defects observed at BoL. The $C_6^-$ depth distribution further confirms significant SEI thickness non-uniformity, with thin SEI regions corresponding to these "pits". This is corroborated by the darker gray curve (Low EI – EoL) in Fig. 6d, showing slightly higher intensity than the light orange curve (High EI – EoL) within the 0–60 s sputter time range, indicating thinner SEI coverage in these regions, while other regions remain covered by a thicker SEI layer. Examining the organic and inorganic

SEI components via $CH_3O^-$ and $LiO^-$ fragments, the low EI group shows a notably higher $t_{stable}$ for $CH_3O^-$, indicating a thicker organic SEI layer. In contrast, the $t_{stable}$ for $LiO^-$ is only slightly higher. Still, its intensity within the 0–40 s sputter range is significantly greater, suggesting earlier exposure of the inorganic layer, likely linked to surface defects.

Based on the NI, SEM, and TOF-SIMS observations, the differing aging trends of the two initial EI groups are summarized in Fig. 6g. For the high EI group, electrolyte depletion during aging is replenished by excess EI from the gas bag region, limiting aging stage to ExEIL. The graphite surface remains fully wetted, maintaining uniform local current density and leading to homogeneous SEI growth and a slower degradation rate in capacity fading stages I-III. In contrast, the low EI group experiences electrode region electrolyte depletion without effective replenishment, progressing to PDEIL aging. Pore drying at both the macro (different regions of the entire electrode, cm-scale heterogeneity in Fig. 4b) and meso (different areas on the graphite particle surfaces, submicron-scale heterogeneity in Fig. 6c) scales causes uneven electrolyte wetting, leading to significant SEI growth heterogeneity. Increased local current density from pore drying reduces the electrochemically active area, leading to thicker organic and inorganic SEI layers in electrolyte-covered regions. This elevated current density also explains the faster stage II capacity fading observed in the Low EI group compared to the High EI group (Fig. 1a). When interpreting the TOF-SIMS results, it is worth noting that the porous structure of the organic SEI layer may allow primary ions to penetrate quickly in some regions, sputtering secondary ions from the inorganic layer beneath. Therefore, $t_{stable}$ values do not necessarily reflect the internal-external sequence of SEI layers but serve as a reference for thickness variations. So we adopt the prevailing understanding of SEI as a bilayer structure with an organic outer layer and an inorganic inner layer.

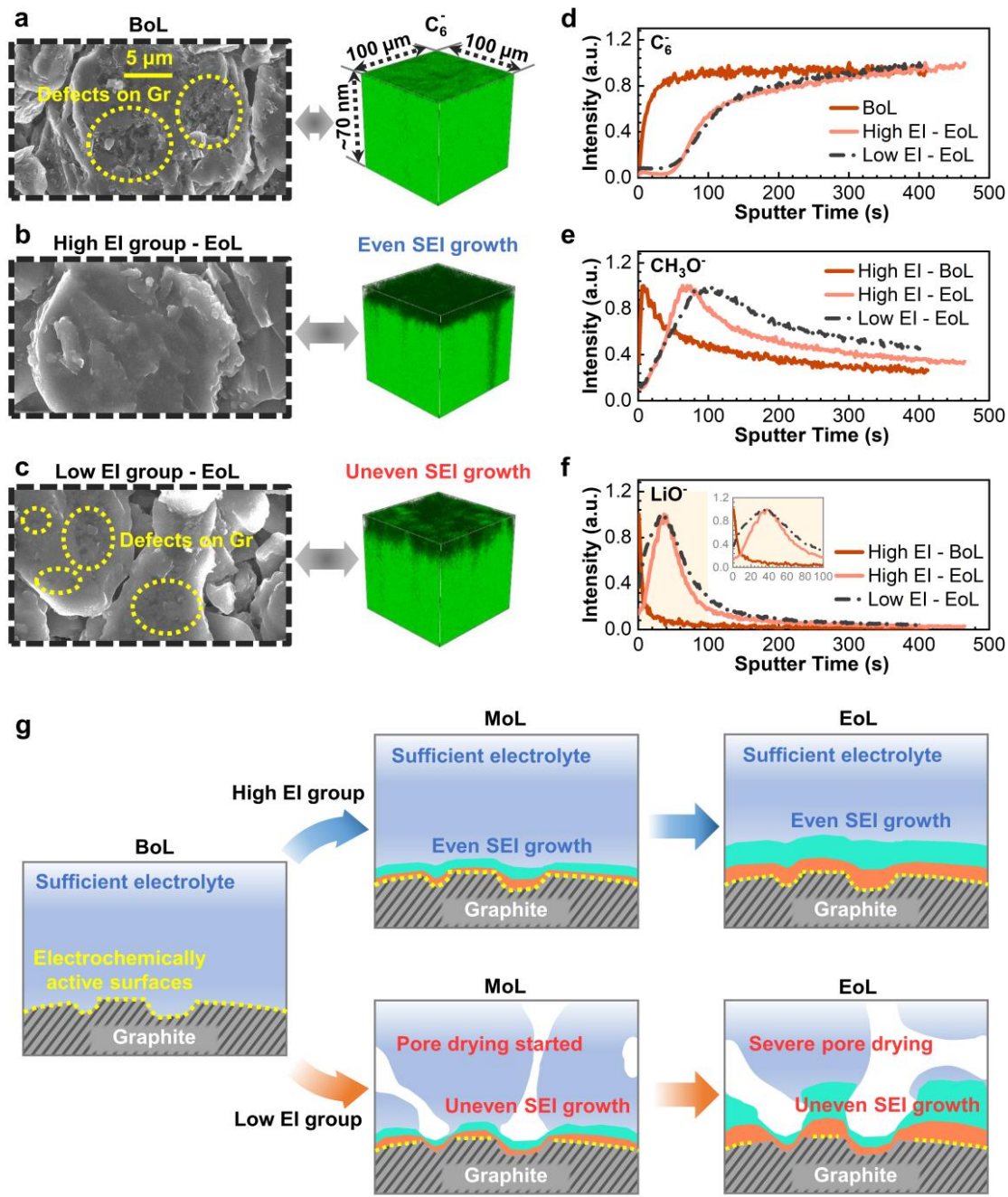

**Fig. 6 | SEI differences after deep aging of two initial EI cells.** a–c, SEM images of the graphite NE at BoL and EoL for two cell groups, accompanied by corresponding 3D rendered depth distributions of $C_6^-$ fragments derived from TOF-SIMS analysis. d–f, TOF-SIMS depth profiles of $C_6^-$, $CH_3O^-$, and $LiO^-$ fragments for the respective samples. g, Schematic illustration of the SEI growth mechanism differences during the aging process of the two initial EI cell groups.

## Conclusion

This study provides a comprehensive understanding of EI dynamics and their impact on the degradation of LIBs. Using large field-of-view NI, electrochemical diagnostics, and post-mortem analyses, we present a non-destructive, quantitative NI-based technique for assessing EI in battery cells and identify two distinct phases of electrolyte degradation: ExEIL and PDEIL, separated by a threshold at 3.18 g Ah$^{-1}$. By evaluating the effects of lithium salt concentration, solvents ratios, cell pressure, and electrode material expansion-contraction cycles on neutron transmission, we demonstrate, for the first time, the use of NI to quantitatively monitor SEI growth during aging. The changes in NI signals due to SEI formation are primarily attributed to the enrichment of hydrogen nuclei in the SEI. The *operando* cyclic aging experiment demonstrated the utility of NI as a non-destructive tool for tracking electrolyte dynamics with sufficient spatial and temporal resolution, enabling detailed observation of electrolyte distribution and evolution throughout the aging process. During the 65°C accelerated aging, high initial EI cells benefit from excess electrolytes in the gas bag, which promotes uniform SEI growth and mitigates capacity fade. In contrast, low initial EI cells transition into PDEIL earlier, leading to pore drying, non-uniform SEI growth, and accelerated degradation. Specifically, the combined use of NI, TOF-SIMS and SEM reveals that high EI cells exhibit more consistent SEI growth, ensuring uniform surface coverage and reducing localized degradation. In contrast, low EI cells experience uneven SEI formation due to macro- and meso-scale dry-out, exacerbating local current density and accelerating capacity loss. This work underscores the importance of optimizing EI design to enhance LIB performance and longevity. The methodology presented is applicable to other energy storage systems and offers valuable insights into improving battery sustainability. Future research should focus on exploring various cell chemistries, the contributions of different SEI components, and the effects of electrolytes under diverse aging conditions, such as fast-charging and low-temperature cycling. This will help refine strategies for developing safer and more durable LIB technologies.

## Methods

### Preparation of cells with two different initial EI

All pouch cells used in this study were 110 mAh laminated cells produced by Contemporary Amperex Technology Co., Limited 21C Innovation Lab (CATL 21C-LAB, Ningde, China). The cells are designed for energy storage applications and consist of two single-side coated LFP positive electrodes (PE), one double-side coated graphite NE, and two separators, forming a PE-Separator-NE-Separator-PE laminated structure. The operational voltage range of the cells is 2.5–3.65 V. Further details regarding the cell design parameters are provided in Fig. S1 and Supplementary Table 1. The cells were received as vacuum-sealed dry cells (electrolyte not yet injected). Commercial LP57 electrolyte with 1% vinyl carbonate (VC) was introduced into the dry cells. Based on the calculated pore volume of the cell design (~225 µL), we confirmed that 300 µL of electrolyte fully saturates the electrode pore structure with 33% excess electrolyte in the gas bag region, while 400 µL results in complete pore saturation with an additional 78% surplus EI. These two initial EI cells allow for studying two distinct aging scenarios: electrolyte depletion with and without subsequent dry-out. It should be noted that the 33% excess electrolyte in the 300 µL group was introduced to account for variations in cell porosity, electrolyte injection volume, distribution heterogeneity, and initial formation-induced electrolyte consumption. This excess ensures that, at BoL, the cell approaches the PDEIL stage without experiencing significant dry-out.

The electrolyte injection and secondary sealing processes were performed in the dry room (Innovation Laboratory for Sciences and Technologies of Energy Materials of Fujian Province (IKKEM), Xiamen, China), where the dew point temperature was maintained below –70°C.

**Electrochemical cycling and diagnostic protocols**

All electrochemical tests were performed under controlled conditions. Direct current (DC) tests were conducted using a Neware CT-5080-5V6A-ATL battery testing system (Neware Technology Limited China), while alternating current (AC) tests employed a BioLogic VMP-3e electrochemical workstation. All test environment temperatures were controlled with LIK CZ-D-2-300D thermal chambers (Guangdong LIK Industrial Co., Ltd., China). Throughout the aging process, RPTs were applied to monitor the thermodynamic and kinetic degradation modes, including Extrap-EMF and Short-Rest-GITT. The DCRs were calculated from Short-Rest-GITT data following:

$$\Delta t\, DCR = \frac{V_0 - V_{\Delta t}}{\Delta I} \qquad (4)$$

where $\Delta t\,DCR$ is the DCR calculated from the time interval $\Delta t$ after the transition point from the current pulse period to the relaxation period. $V_0$ and $V_{\Delta t}$ represent the cell voltage before and $\Delta t$ after the current step, respectively, and $\Delta I$ is the step current.

The RPT design is based on our previous work[25], and the detailed protocols are provided in Supplementary Table 2. Subsequent RPTs were labeled as RPTn, where n denotes the sequence number of the RPT. At BoL (100% SoH) and final degradation states of each cell (90%, 80%, and 70% SoH), TV-EIS@50%SoC was conducted with a BioLogic VMP-3e electrochemical workstation across five temperatures (25, 10, 0, −10, and −20°C) with a frequency range of ~10 mHz to 1 MHz (adjusted slightly for each temperature). There were 10 points per decade in logarithmic spacing and 2 average measures per frequency.

**Neutron imaging experimental setup**

Neutron imaging studies were performed using the Energy-Resolved Neutron Imaging (ERNI) instrument[24] at the China Spallation Neutron Source (CSNS) in Dongguan, China. The field of view was set to maximum aperture size of 20 × 20 cm². A 100 μm thick ZnS scintillator screen was utilized. The distance between the sample back plate and the imaging detector surface is 3 cm. Optical data collection was performed using CCD camera (2048 × 2048 pixels, 15 μm pixel size). The physical resolution was estimated to ~150 μm as determined by the Siemens stars on the imaging resolution test object (see Supplementary Fig. 4). A single image was captured with an exposure time of 20 s. Including the processing and storage time of the imaging software, the effective acquisition time was approx. 40 s. The pouch cell was clamped in an aluminum fixture with flexible polytetrafluoroethylene (PTFE) gaskets under a pressure of 1.5 MPa during the experiment to replicate realistic battery operating conditions.

***Post mortem* analysis**

The quantification of all electrolyte compositions was achieved by the combined analysis of gas chromatography mass spectrometry (GC-MS, Agilent 8860 gas chromatograph (HP-5MS, 30m, 0.25mm inner diameter column) coupled to an Agilent 5977B single quadrupole mass spectrometer) and nuclear magnetic resonance (NMR, Bruker Avance Neo 500 MHz

Spectrometer) reported in our previous work[16]. The cell was cut open and placed into a plastic bottle containing approximately 90 mL of DMC for complete electrolyte extraction. The bottle was then sealed and shaken thoroughly to ensure full dissolution of the electrolyte. Afterward, the solution was filtered, and separate aliquots were prepared - each spiked with internal standards - for GC-MS and NMR analyses. Specifically, solvents and additives (EMC, EC, VC) are quantified by GC-MS and lithium anions ($PF_6^-$) by solution $^{19}F$ NMR. The differential weight method was used to quantify the actual EI of the cell samples. After extracting the electrolyte, all remaining cell components were collected in a glass beaker and placed in a 100 °C oven for over 24 hours. The total electrolyte mass was then determined by subtracting the fully dried mass of these components from their pre-disassembly mass.

The morphology of samples was examined using field emission scanning electron microscopy (FESEM, ZEISS, SUPRA55 SAPPHIRE). Time-of-flight secondary ion mass spectrometry (TOF-SIMS) was performed using an IONTOF M6 instrument (ION-TOF GmbH., Münster, Germany), where each second of sputtering time corresponds to a sputtering depth of 0.2 nm for $SiO_2$. The 3D TOF-SIMS data was acquired over an area of 100 × 100 um using a $Bi^+$ primary ion beam. Given the sensitivity of SEI on the anode to air, the samples for TOF-SIMS testing were transferred directly from the glovebox to the instrument using a specially designed vacuum transfer device.

**Acknowledgements**

This work was supported by the National Key R&D Program of China (grant no. 2021YFB2401800). The authors acknowledge the beamtime at the energy-resolved neutron imaging instrument (ERNI, https://csns.cn/31113.02.CSNS. ERNI) granted from China Spallation Neutron Source (CSNS, https://csns.cn/31113.02.CSNS) and thank the staff members of ERNI for providing technical support and assistance in data collection and analysis. Dr. Lufeng Yang and Prof. Jie Chen acknowledge the support from the funds for science and technology innovation of the Institute of High Energy Physics (E35454U210), Chinese Academy of Sciences (CAS). The authors acknowledge the CATL 21C innovation lab for providing dry cells with excellent consistency.

## Author contributions

Y. Yang acquired project funding and directed the project. J. Chen arranged and provided beamtime at ERNI. Y. Hu and Y. Liao, with input from Y. Yang, planned the project. Y. Liao prepared all cell samples, and together with Y. Hu designed and conducted all electrochemical tests. Y. Hu analyzed all electrochemical data, drawing on TLM insights from J. Lin and using Short-Rest-GITT data-processing software developed by J. Liang and Y. Wei. Y. Hu and Y. Liao designed the NI experiment, while L. Yang, Y. Peng, and S. Tang provided critical advice on its design. Y. Hu, Y. Liao, and L. Yang performed NI experiments with *in situ* monitoring support from K. Zhang, M. Ding, and J. Wu and analyzed the NI data with suggestions from S. Wang. Y. Hu conducted post-mortem analyses with electrolyte quantification assistance from K. Zhang. Y. Hu wrote the manuscript. Y. Yang, J. Chen, A. S., Z. Gong, and Y. Jin supervised the research and contributed to review and editing the manuscript.

## Competing interests

The authors declare no competing interests.